\documentclass[10pt,twocolumn,aps,showpacs]{revtex4}

\begin{document}

\title{Quantum computers can search rapidly by using almost any selective transformations}  
\author{Avatar Tulsi\\
        {\small Department of Physics, Indian Institute of Science, Bangalore-560012, India}}
\email{tulsi9@gmail.com}   
 
\begin{abstract}
The search problem is to find a state satisfying certain properties out of a given set. Grover's algorithm drives a quantum computer from a prepared initial state to the target state and solves the problem quadratically faster than a classical computer. The algorithm uses selective transformations to distinguish the initial state and target state from other states. It does not succeed unless the selective transformations are very close to phase-inversions. Here we show a way to go beyond this limitation. An important application lies in quantum error-correction, where the errors can cause the selective transformations to deviate from phase-inversions. The algorithms presented here are robust to errors as long as the errors are reproducible and reversible. This particular class of systematic errors arise often from imperfections in apparatus setup. Hence our algorithms offer a significant flexibility in the physical implementation of quantum search.
\end{abstract}

\maketitle

\section{Introduction}

Suppose we have a set of $N$ items, $j=0,1,2,\ldots N-1$, and a binary function $f(j)$ which is $1$ if $j$ satisfies certain properties (e.g. if it is a solution to a certain computational problem) and $0$ otherwise. Let $T$ be the set of $M$ items for which $f(j)=1$, i.e. $T=\{j|f(j)=1\}$ and $|T|=M$. Consider the situation when the items are not sorted according to any property, but $f(j)$ can be computed by querying an oracle that outputs $f(j)$ for any input $j$. The search problem is to find an element of $T$ (i.e. a solution) using the minimum number of oracle queries. The best classical algorithm for this problem is to randomly pick an item $j$, use an oracle query to check whether $j\in T$, and then repeating the process till a solution is found. On the average, it takes $O(N/M)$ oracle queries to succeed, since $M/N$ is the probability of the picked item to be a solution.

In a quantum setting, Grover's search algorithm~\cite{grover1} provides a much faster way. The $N$ items are encoded as basis states $|j\rangle$ of an $N$-dimensional Hilbert space, which can be realized using $n=\log_{2}N$ qubits (without loss of generality, we assume $N$ to be a power of $2$). The initial unbiased state is chosen as the equal superposition state, $(1/\sqrt{N})\sum_{j}|j\rangle$, generated by applying the Walsh-Hadamard transformation $W$ on $|0\rangle$. The target state $|t\rangle$ can be any normalised state $\sum_{j\in T}a_{j}|j\rangle$ within the target subspace, since measuring $|t\rangle$ will always give a solution. Grover's algorithm obtains $|t\rangle$ by applying $O(\sqrt{N/M})$ iterations of the operator $\mathcal{G}=WI_{0}WI_{t}$ on $W|0\rangle$. Here $I_{t}=\sum_{j}(-1)^{\delta_{jt}}|j\rangle\langle j|$ and $I_{0}=\sum_{j}(-1)^{\delta_{j0}}|j\rangle\langle j|$ are the selective phase-inversions of $|t\rangle$ and $|0\rangle$ states respectively. Grover's algorithm thus provides a quadratic speedup over the classical algorithm, as each iteration of $\mathcal{G}$ uses one oracle query to implement $I_{t}$.

Grover showed that his algorithm works even if the Walsh-Hadamard transform $W$ is replaced by \emph{almost} any unitary operator $U$~\cite{grover2}. In this case, the initial state $U|0\rangle$ is a general (not necessarily equal) superposition of the basis states. The operator $\overline{\mathcal G}=UI_{0}U^{\dagger}I_{t}$ is iteratively applied to $U|0\rangle$, and the target state $|t\rangle$ is obtained after $O(1/\alpha_{U})$ iterations, where $\alpha_{U}=\sqrt{\sum_{j\in T}|U_{j0}|^{2}}$ is the projection of $U|0\rangle$ on the target subspace (for $U=W$, $\alpha_{W}=\sqrt{M/N}$). As the probability of getting a target state upon measuring $U|0\rangle$ is $\alpha_{U}^{2}$, the target state can be obtained classically by $O(1/\alpha_{U}^{2})$ preparations of $U|0\rangle$ and subsequent projective measurements. Hence, the quantum algorithm provides a quadratic speedup over this simple scheme by doing the same job in $O(1/\alpha_{U})$ steps. This generalization is known as quantum amplitude amplification~\cite{grover2,qaa}, and forms the backbone of many other quantum algorithms. It has an important application when in a physical implementation $W$ gets replaced by $U$ due to some unavoidable error. The algorithm succeeds as long as $U$ and $U^{\dagger}$ can be consistently implemented even when we do not know their precise form, making it intrinsically robust against certain types of errors. In the case of quantum search, provided $\alpha_{U}\not\ll\alpha_{W}$, there is not much of a slowdown and hence almost any transformation is good enough.

Quantum amplitude amplification often fails, however, when the selective phase-inversions $\{I_{t}, I_{0}\}$ are replaced by other selective transformations, say $\{S_{t},S_{0}\}$. Consider the simple case when $S_{t}=R_{t}^{\phi}=\sum_{j}e^{i\phi\delta_{jt}}|j\rangle\langle j|$ and $S_{0}=R_{0}^{\varphi}=\sum_{j}e^{i\varphi\delta_{j0}}|j\rangle\langle j|$ are the selective phase rotations of $|t\rangle$ and $|0\rangle$ states by angles $\phi$ and $\varphi$ respectively. The well-known phase matching condition~\cite{long1,hoyer1} demands $|\phi-\varphi| \ll \alpha_{U}$ for quantum amplitude amplification to succeed. This is a very strict condition for $\alpha_{U} \ll 1$, while the quadratic speedup is not of much use for large $\alpha_{U}$. In fact, systematic phase mismatching (i.e. $|\phi-\varphi| \not\ll \alpha_{U}$) is known to be the dominant gate imperfection in implementing quantum amplitude amplification, posing an intrinsic limitation to the size of database that can be searched~\cite{long2}.

In this work, we show that a successful quantum search can be obtained with almost any selective transformations $\{S_{t},S_{0}\}$, provided their inverse transformations $\{S_{t}^{\dagger},S_{0}^{\dagger}\}$ are also available. This is useful in situations where the errors are \emph{reproducible} (i.e. every time we ask for the transformation $\mathcal{A}$ the system implements the transformation $\mathcal{B}$) as well as \emph{reversible} (i.e. whenever we ask for the transformation $\mathcal{A}^{\dagger}$ the system implements the transformation $\mathcal{B}^{\dagger}$). For instance, such systematic errors arise when there is incorrect calibration of the instrumentation. In the following, we present two algorithms in this category, one iterative and the other recursive.

In section II, we consider the case of diagonal selective transformations, which rotate the phases of the desired states by \emph{any} amount (unlike the selective phase-inversions that change the phase by $\pi$) but leave all the other (non-desired) states unchanged. We then construct an operator which yields a successful quantum search algorithm when iterated on the initial state, and we show the algorithm to be optimal up to a constant factor. This iteratve algorithm does not work in the case when diagonal selective transformations also perturb the non-desired states. In section III, we design a recursive quantum search algorithm for such transformations provided they are not too far off from the selective phase-inversions. The algorithm requires $O(1/\alpha_{U}^{1+O(\Delta_{t}^{2},\Delta_{0}^{2})})$ queries, where $\Delta_{t}=\|S_{t}-I_{t}\|$ and $\Delta_{0}=\|S_{0}-I_{0}\|$ are the distances of selective transformations from the corresponding selective phase-inversions, assumed to be small.
It is straightforward to extend the above two algorithms to situtations where the selective transformations are non-diagonal. We describe that in section IV, together with possible applications of our algorithms to quantum error correction, quantum workspace errors and bounded-error quantum search.

\section{Iterative algorithm}

Consider those selective transformations $\{S_{t},S_{0}\}$ which rotate the phases of the desired states by arbitrary angles but leave all the other states unchanged. In case of $|0\rangle$, there is only one desired state and $S_{0}=R_{0}^{\varphi}=I-(1-e^{i\varphi})|0\rangle\langle 0|$. In case of $|t\rangle$, there can be multiple target states and the rotation phase can be different for different target states, so $S_{t}=R_{t}=\sum_{j}e^{i\phi_{j}\delta_{jt}}|j\rangle \langle j|$. If we iteratively apply the generalized quantum amplitude amplification operator $\widetilde{\mathcal G}=UR_{0}^{\varphi}U^{\dagger}R_{t}$ on the initial state $U|0\rangle$, we will not succeed in getting a target state unless the phase matching condition is satisfied.
 
Instead, we iteratively apply a different operator, $\mathcal{T} = UR_{0}^{-\varphi}U^{\dagger}R_{t}^{\dagger}UR_{0}^{\varphi}U^{\dagger}R_{t}$, on the initial state $U|0\rangle$. It uses two oracle queries, one for $R_{t}$ and another for $R_{t}^{\dagger}$. It also uses $R_{0}^{\varphi\dagger}=R_{0}^{-\varphi}$ along with $R_{0}^{\varphi}$. Thus, unlike $\widetilde{\mathcal G}$, it makes explicit use of the inverse transformations $\{R_{t}^{\dagger},R_{0}^{\dagger}\}$. Observe that $\mathcal{T}$ is a product of two selective phase rotations: $UR_{0}^{-\varphi}U^{\dagger}$ is a rotation by $-\varphi$ of the state $U|0\rangle$, and $R_{t}^{\dagger}UR_{0}^{\varphi}U^{\dagger}R_{t}$ is a rotation by $\varphi$ of the state $R_{t}^{\dagger}U|0\rangle$. We therefore have
\begin{equation}
\mathcal{T}=UR_{0}^{-\varphi}U^{\dagger}R_{\sigma}^{\varphi},~~
|\sigma\rangle \equiv R_{t}^{\dagger}U|0\rangle. \label{Tfirstdefine}
\end{equation}

Let $|\tau\rangle$ be a state orthogonal to $|\sigma\rangle$ in the two-dimensional subspace spanned by $U|0\rangle$ and $|\sigma\rangle$, such that up to an overall phase
\begin{equation}
U|0\rangle=\cos\theta|\sigma\rangle+\sin\theta|\tau\rangle
=\cos\theta R_{t}^{\dagger}U|0\rangle+\sin\theta|\tau\rangle.
\label{U0expansion}
\end{equation}
For a general vector $|\psi(a,b)\rangle=a |\sigma\rangle+b |\tau\rangle$ in this subspace, we have
\begin{equation}
\mathcal{T}|\psi(a,b)\rangle=UR_{0}^{-\varphi}U^{\dagger}R_{\sigma}^{\varphi}|\psi(a,b)\rangle = UR_{0}^{-\varphi}U^{\dagger}|\psi(a e^{i\varphi},b)\rangle. \label{subspacepreserve1}
\end{equation} 
As $UR_{0}^{-\varphi}U^{\dagger}|\psi\rangle=|\psi\rangle-zU|0\rangle$ with $z = (1-e^{-i\varphi})\langle 0|U|\psi\rangle$, we have 
\begin{equation}
\mathcal{T}|\psi(a,b)\rangle=|\psi(a e^{i\varphi}-z\cos\theta,b-z\sin\theta)\rangle. \label{subspacepreserve2}
\end{equation} 
Hence, $\mathcal{T}$ preserves this two-dimensional subspace. For any vector within this subspace, we can also write $R_{\sigma}^{\varphi}=e^{i\varphi}R_{\tau}^{-\varphi}$, and so $\mathcal{T}$ is equivalent to $UR_{0}^{-\varphi}U^{\dagger}R_{\tau}^{-\varphi}$ up to an overall phase, i.e.
\begin{equation}
\mathcal{T} \cong UR_{0}^{-\varphi}U^{\dagger}R_{\tau}^{-\varphi} \label{Tseconddefine}
\end{equation}

The above operator is a special case of the generalized quantum amplitude amplification operator with $|\tau\rangle$ as the effective target state. It satisfies the phase matching condition by construction. One may wonder that the phase matching condition is not satisfied in the form $\mathcal{T}=UR_{0}^{-\varphi}U^{\dagger}R_{\sigma}^{\varphi}$ as $\varphi \neq -\varphi$ in general. But the phase matching condition was derived assuming $\alpha_{U}\ll 1$, and it cannot be used with $|\sigma\rangle$ as the effective target state because $\langle\sigma|U|0\rangle=\langle 0|U^{\dagger}R_{t}U|0\rangle$ is close to $1$. That is why we converted $R_{\sigma}^{\varphi}$ to $R_{\tau}^{-\varphi}$.

Now applying $\mathcal{T}$ on the initial state $U|0\rangle$ rotates it towards the state $|\tau\rangle$ by an angle $2\theta\sin\frac{\varphi}{2}$~\cite{long3}. After $n$ iterations of $\mathcal{T}$, we get
\begin{equation}
\mathcal{T}^{n}U|0\rangle=\cos\theta_{n}|\sigma\rangle+\sin\theta_{n}|\tau\rangle,~
\theta_{n}=\theta\left(1+2n\sin\frac{\varphi}{2}\right). \label{TnU0} 
\end{equation}
For $n=\lfloor\pi/(4\theta\sin\frac{\varphi}{2})\rfloor$, $\theta_{n}$ is close to $\pi/2$ and $\mathcal{T}^{n}U|0\rangle$ is close to $|\tau\rangle$. Further iterations of $\mathcal{T}$ rotate the state away from $|\tau\rangle$, displaying a cyclic motion in the two-dimensional subspace as in case of Grover's algorithm.

To understand the significance of the state $|\tau\rangle$, we use the expansions $U|0\rangle=\sum_{j}U_{j0}|j\rangle$ and $R_{t}^{\dagger}U|0\rangle = \sum_{j}U_{j0}e^{-i\phi_{j}\delta_{jt}}|j\rangle$ in Eq. (\ref{U0expansion}), and obtain
\begin{equation}
|\tau\rangle = \frac{1}{\sin\theta}\sum_{j}U_{j0}\left(1-\cos\theta e^{-i\phi_{j}\delta_{jt}}\right)|j\rangle.
\label{tauexpansion}
\end{equation}
Here $\cos\theta =|\langle 0|U^{\dagger}R_{t}U|0\rangle| =\left|\sum_{j}|U_{j0}|^{2}e^{i\phi_{j}\delta_{jt}}\right|$, and since $\sum_{j\in T}|U_{j0}|^{2}=\alpha_{U}^{2}$, we have the bound $\cos\theta \geq 1-2\alpha_{U}^{2}$ or $|\theta| \leq 2\alpha_{U}$. Hence, $\langle j|\tau\rangle=U_{j0}O(\alpha_{U})$ for $j\not\in T$, and the projection of $|\tau\rangle$ on the non-target subspace is $\sqrt{\sum_{j\not\in T}|\langle j|\tau\rangle|^{2}}=O(\alpha_{U})$. This projection is very small, which makes $|\tau\rangle$ almost a state in the target subspace, $|\langle t|\tau\rangle|=1-O(\alpha_{U}^{2})\approx 1$. 

The number of queries needed to reach the state $|\tau\rangle$ is twice the number of iterations of $\mathcal{T}$, as each iteration uses two queries. We therefore have $Q = \pi/(2\theta\sin\frac{\varphi}{2})$. The normalization condition for Eq. (\ref{tauexpansion}), $|\langle \tau|\tau\rangle|=1$, gives
\begin{equation}
\sin^{2}\theta = (1-\cos\theta)^{2} + \sum_{j\in T}4|U_{j0}|^{2}\cos\theta\sin^{2}\frac{\phi_{j}}{2}.
\end{equation}
For small $\theta$, this yields $\theta=\sqrt{\sum_{j\in T}4|U_{j0}|^{2}sin^{2}\frac{\phi_{j}}{2}}$, and
\begin{equation}
Q=\frac{\pi}{4\sin\frac{\varphi}{2}\sqrt{\sum_{j\in T}|U_{j0}|^{2}\sin^{2}\frac{\phi_{j}}{2}}}.
\end{equation}

For later reference, we point out that the state $|\tau\rangle$ is close to target state only because $\theta=O(\alpha_{U})$. That is true for the $R_{t}$ transformations which act only within the target subspace, but may not be true for general selective transformations $S_{t}$ which perturb non-target states also. More generally, Eq. (\ref{tauexpansion}) provides $|\langle t|\tau\rangle|=O(\frac{\alpha_{U}}{\theta})$, and the iterative algorithm can amplify the projection on the target subspace by a maximum factor of $1/\theta$. That may be too small for a general selective transformation to reach a target state. In section III, we use the idea of recursion to overcome this limitation.

\subsection{Comparison with Grover's algorithm}

When $\{S_{t},S_{0}\}=\{I_{t},I_{0}\}$, i.e. when $\varphi=\phi_{j}=\pi$, the operator $\mathcal{T}$ is simply $2$ steps of the quantum amplitude amplification algorithm. To demonstrate the difference between $\mathcal{T}$ and $\widetilde{\mathcal G}^{2}$ for general $\{S_{t},S_{0}\}$, consider the situation where $S_{t}=R_{t}^{\phi}$, i.e. rotation angle $\phi$ is the same for all target states. Then $\mathcal{T}=UR_{0}^{-\varphi}U^{\dagger}R_{t}^{-\phi}UR_{0}^{\varphi}U^{\dagger}R_{t}^{\phi}$ while $\widetilde{\mathcal G}^{2}=UR_{0}^{\varphi}U^{\dagger}R_{t}^{\phi}UR_{0}^{\varphi}U^{\dagger}R_{t}^{\phi}$. The quantum amplitude amplification algorithm succeeds in this case only if the phase-matching condition is satisfied, $|\phi-\varphi|\ll \alpha_{U}$~\cite{long1}. On the other hand, there is no such restriction on our algorithm which succeeds using $\pi/(4\alpha_{U}\sin\frac{\varphi}{2}\sin\frac{\phi}{2})$ queries. As Grover's optimal algorithm takes $\pi/(4\alpha_{U})$ queries, the slowdown is only by the constant factor $1/(\sin\frac{\varphi}{2}\sin\frac{\phi}{2})$. As long as $\phi,\varphi$ are not very small, not much is lost, and hence almost any selective transformation can be used for quantum search. 

This particular case has been experimentally verified on an NMR quantum information processor~\cite{avik}, which compares the performances of Grover's and our algorithm for small $\alpha_{U}$ and $\phi\neq\varphi$. The experimental data confirms the theoretical prediction that our algorithm succeeds in getting the target state while Grover's algorithm does not. 
In a more general case, the rotation angle $\phi_{j}$ can be different for different target states and $S_{t}=R_{t}=\sum_{j}e^{i\phi_{j}\delta_{jt}}|j\rangle \langle j|$. It is shown in an upcoming paper~\cite{tulsi} that in this case, iterating the operator $\widetilde{\mathcal{G}} = UR_{0}^{\varphi}U^{\dagger}R_{t}$ amplifies only those target states which satisfy the phase-matching condition i.e. for which $|\phi_{j}-\varphi|\ll |U_{j0}|$. (If there are no such target states then iterating $\widetilde{\mathcal{G}}$ will not succeed in getting a target state.) The target state is obtained after $O(1/\alpha_{U}')$ iterations where $\alpha_{U}' = \sqrt{\sum_{j:|\phi_{j}-\varphi|\ll |U_{j0}|}|U_{j0}|^{2}} \leq \alpha_{U}$, and the algorithm suffers a slowdown by a factor $\alpha_{U}/\alpha_{U}'$. There is no such restriction on our algorithm and the full amplitude along the target states can be utilised, irrespective of any phase-matching condition.

Quantum amplitude amplification is often described as a rotation in the two-dimensional space spanned by the initial state $U|0\rangle$ and the target state $|t\rangle$. Here, we have provided a new insight suggesting that it is better to interpret quantum search as a rotation in the two-dimensional space spanned by the initial state $U|0\rangle$ and the oracle-modified initial state $R_{t}^{\dagger}U|0\rangle$. $\mathcal{T}$ is then the fundamental unit of quantum search rather than $\widetilde{\mathcal G}$. The advantage of the operator $\mathcal{T}$ is that it uses the selective transformations and their inverses in such a way that the phase-matching condition is \emph{effectively} satisfied to produce a successful quantum search.

\subsection{New search Hamiltonian}

Grover's algorithm is a digital algorithm in the sense that it uses a discrete set of unitary operators and applies them sequentially on the initial state to reach the target state. Farhi and Gutmann developed an analog version of the algorithm~\cite{farhi}, which shows that any initial state, when evolved under a particular search Hamiltonian for a certain amount of time, will evolve to the target state. Their search Hamiltonian is given by $\mathcal{H}_{FG}=\mathcal{H}_{U|0\rangle}+\mathcal{H}_{|t\rangle}$, where $\mathcal{H}_{U|0\rangle}=I-U|0\rangle \langle 0|U^{\dagger}$ and $\mathcal{H}_{|t\rangle}=I-|t\rangle \langle t|$ are projector Hamiltonians. More general search Hamiltonians have been presented subsequently by Fenner~\cite{fenner}, and by Bae and Kwon~\cite{baekwon}.  

The algorithm developed above suggests a new search Hamiltonian
\begin{equation} 
\mathcal{H}_{\rm new}=\mathcal{H}_{U|0\rangle}+\mathcal{H}_{R_{t}^{\dagger}U|0\rangle}=\mathcal{H}_{U|0\rangle}+R_{t}^{\dagger}\mathcal{H}_{U|0\rangle}R_{t}.
\end{equation}
The second term of $\mathcal{H}_{\rm new}$ is just the first term but in a basis rotated by the oracle transformation $R_{t}$. $\mathcal{H}_{\rm new}$ can be analysed the same way as was done by Farhi and Gutmann, in the two-dimensional subspace spanned by $U|0\rangle$ and $R_{t}^{\dagger}U|0\rangle$. When evolved using $\mathcal{H}_{\rm new}$ for a certain amount of time, the initial state becomes the state $|\tau\rangle$, which is very close to a target state as shown. 

$\mathcal{H}_{\rm new}$ has certain physical implementation advantages over $\mathcal{H}_{FG}$. Consider the situation when implementation errors perturb $\mathcal{H}_{FG}$ to $(1-s)\mathcal{H}_{U|0\rangle}+(1+s)\mathcal{H}_{|t\rangle}$, i.e. one term is enhanced while the other gets reduced. Analysing this perturbed Hamiltonian as is done in Ref.~\cite{roland}, it is easy to see that one reaches a target state only if $|s|<O(\alpha_{U})$. This is analogous to the phase-matching condition, and as $\alpha_{U}\ll 1$, it is a strict condition. There is no such restriction, however, on the new search Hamiltonian as it is the sum of the same term in two different bases. For example, calibration errors remain \emph{effectively} the same for both terms, making $\mathcal{H}_{\rm new}$ robust with $s=0$.

\section{Recursive algorithm}

In this section, we consider those diagonal selective transformations $\{S_{t},S_{0}\}$ which may also perturb the non-desired states, unlike the transformations $\{R_{t},R_{0}\}$ discussed in previous section which leave them unperturbed. We assume the perturbations to be small, i.e. $\|S_{t}-I_{t}\|=\Delta_{t}$ and $\|S_{0}-I_{0}\|=\Delta_{0}$ where $\Delta_{t}$ and $\Delta_{0}$ are small. More explicitly, 
\begin{eqnarray}
S_{t}=\sum_{j}e^{i\phi_{j}}|j\rangle \langle j|\ ,\ \phi_{j}=\pi\delta_{jt}+\epsilon_{j}\ ,\ |\epsilon_{j}|\leq \Delta_{t}; \nonumber \\
S_{0}=\sum_{j}e^{i\varphi_{j}}|j\rangle \langle j|\ ,\ \varphi_{j}=\pi\delta_{j0}+\mu_{j}\ ,\ |\mu_{j}|\leq \Delta_{0}. \label{S0Stexpansion}
\end{eqnarray}

For such transformations, the iteration of operator $US_{0}U^{\dagger}S_{t}$ on the initial state $U|0\rangle$ may not give us a target state. As $O(1/\alpha_{U})$ iterations of $UI_{0}U^{\dagger}I_{t}$ on $U|0\rangle$ gives us a target state, it is easy to see that as long as $\{\Delta_{t},\Delta_{0}\}<O(\alpha_{U})$, iterating the operator $US_{0}U^{\dagger}S_{t}$ on $U|0\rangle$ will also bring us close to target state. But when $\{\Delta_{t},\Delta_{0}\}\geq O(\alpha_{U})$, the phase-matching condition required for a successful quantum search may not be satisfied. Another way to see this is to analyse the eigenspectrum of $US_{0}U^{\dagger}S_{t}$. Its distance from $UI_{0}U^{\dagger}I_{t}$ is $O(\Delta_{t},\Delta_{0})$. The two eigenvalues of $UI_{0}U^{\dagger}I_{t}$, relevant for quantum search, are separated by $O(\alpha_{U})$. A perturbation greater than $O(\alpha_{U})$ will in general shift them too much to maintain a successful quantum search. Note again that we are considering $\alpha_{U}\ll 1$, which makes the iterative quantum amplitude amplification very sensitive to small errors.

The recursive quantum search algorithm is defined, at the $m^{th}$ level, by the relation
\begin{equation}
U_{m}|0\rangle = U_{m-1}S_{0}U_{m-1}^{\dagger}S_{t}U_{m-1}|0\rangle,
\end{equation}
with $U_{0}\equiv U$. At the first level, $U_{1}|0\rangle=(US_{0}U^{\dagger}S_{t})U|0\rangle$ is a simple generalization of the quantum amplitude amplification step $\overline{\mathcal G}U|0\rangle$. But at higher levels, the operators $U_{m}$ involve $\{S_{t},S_{0}\}$ as well as $\{S_{t}^{\dagger},S_{0}^{\dagger}\}$, and cannot be expressed using repeated iterations of a single operator like $\overline{\mathcal G}$. (For instance, $U_{2} = U_{1}S_{0}U_{1}^{\dagger}S_{t}U_{1} = US_{0}U^{\dagger}S_{t}US_{0}U^{\dagger}S_{t}^{\dagger}US_{0}^{\dagger}U^{\dagger}S_{t}US_{0}U^{\dagger}S_{t}U$ involves operators $S$ and $S^{\dagger}$ in a non-periodic pattern.) The idea of recursive quantum search is not new. It has been used by Hoyer {\it et al.}~\cite{hoyerbound} and by Grover~\cite{private} for specific error models, as discussed in the next section. What is new here is the demonstration that recursion works even for general errors. 

The number of queries used at the $m^{th}$ level of recursion is determined by the relation $q_{m}=3q_{m-1}+1$, since $q_{m-1}$ is the number of queries used by $U_{m-1}$ and $S_{t}$ needs one extra query. Using the fact that $q_{0}=0$ (as implementing $U$ does not need any query), we get
\begin{equation}
q_{m}=\frac{3^{m}-1}{2}=\Theta(3^{m}).
\end{equation}
The recursive algorithm increases the number of queries in a geometric progression with the level number, a factor of $3$ in the present case. On the other hand, the iterative algorithm increases the number of queries in an arithmetic progression with the iteration number, a step of $2$ in the algorithm of the previous section. We will see that the larger jumps in the allowed number of queries for the recursive algorithm are not a major disadvantage, because the total number of queries needed to obtain the target state remains about the same. (The worst case overhead is a tolerable factor of $3$ in the number of queries.) 

At the first level, the initial state $U|0\rangle$ evolves to $U_{1}|0\rangle$, whose projection on the target subspace is $\alpha_{U_{1}}=\sqrt{\sum_{j\in T}|(U_{1})_{j0}|^{2}}$. In recursive quantum search, what matters is the amplification factor 
\begin{equation}
\kappa =\frac{\alpha_{U_{1}}}{\alpha_{U}} = \sqrt{\frac{\sum_{j\in T}|(U_{1})_{j0}|^{2}}{\sum_{j\in T}|U_{j0}|^{2}}}, \label{kappaequation}
\end{equation}
and the target state can be obtained using $O(1/\alpha_{U}^{\log_{\kappa}3})$ queries. To get the nearly optimal algorithm, the amplification factor $\kappa$ should be as close to $3$ (the number of $U$-type operators used by $U_{1}$) as possible. We will show that for small $\{\Delta_{t},\Delta_{0}\}$, $\kappa$ is indeed close to $3$, and the performance of recursive algorithm is close to the optimal algorithm that takes $O(1/\alpha_{U})$ queries.

We estimate $\kappa$ by estimating the ratio $\rho_{j} = |(U_{1})_{j0}/U_{j0}|$ for $j\in T$. In terms of $\rho_{j}$, we have
\begin{equation}
\kappa = \sqrt{\frac{\sum_{j\in T}\rho_{j}^{2}|U_{j0}|^{2}}{\sum_{j\in T}|U_{j0}|^{2}}}. \label{kapparho}
\end{equation} 
Clearly if $\rho_{j}$ is close to $3$ for each $j\in T$, then $\kappa$ is also close to $3$. To find $\rho_{j}$, let
\begin{equation}
|\psi\rangle = S_{t}U|0\rangle = \sum_{j}U_{j0}e^{i\phi_{j}}|j\rangle, \label{psidefinition}
\end{equation}
so that $U_{1}|0\rangle = US_{0}U^{\dagger}|\psi\rangle$. We decompose $S_{0}$ as $S_{0}=S_{0}'\cdot R_{0}^{\varphi_{0}}$, where $S_{0}'=|0\rangle\langle 0|+\sum_{j\neq 0}e^{i\varphi_{j}}|j\rangle \langle j|$ leaves the $|0\rangle$ state unchanged but acts like $S_{0}$ on all the other states, and $R_{0}^{\varphi_{0}}$ is a selective phase-rotation of the $|0\rangle$ state. We have $U_{1}|0\rangle=US_{0}'U^{\dagger}UR_{0}^{\varphi_{0}}U^{\dagger}|\psi\rangle$. With $UR_{0}^{\varphi_{0}}U^{\dagger}|\psi\rangle =|\psi\rangle-(1-e^{i\varphi_{0}})\langle 0|U^{\dagger}|\psi\rangle U|0\rangle$ and $1-e^{i\varphi_{0}}= 2e^{i\mu_{0}/2}\cos\frac{\mu_{0}}{2}$, we get 
\begin{equation}
U_{1}|0\rangle = US_{0}'U^{\dagger}|\psi'\rangle, \label{U10}
\end{equation} 
where
\begin{equation}
|\psi'\rangle=\sum_{j}U_{j0}\left(e^{i\phi_{j}}-2e^{i\mu_{0}/2}\cos\frac{\mu_{0}}{2}\beta\right)|j\rangle. \label{psiprimefirst}
\end{equation}
Here $\beta = \langle 0|U^{\dagger}|\psi\rangle=\sum_{j}|U_{j0}|^{2}e^{i\phi_{j}}$. As $\phi_{j}=\pi\delta_{jt}+\epsilon_{j}$, the bound $|\epsilon_{j}|\leq \Delta_{t}$ gives 
\begin{equation}
(1- Re(\beta)) \leq 0.5\Delta_{t}^{2}+2\alpha_{U}^{2}\ ,\ |Im(\beta)|\leq \Delta_{t}. \label{betanonprimebound}
\end{equation} 
Since $\Delta_{t}^{2}$ and $\alpha_{U}^{2}$ are small, we can write $\beta=|\beta|e^{i\xi}$, where $|\xi| \leq \Delta_{t}$. Then
\begin{equation}
|\psi'\rangle = \sum_{j}U_{j0}e^{i\phi_{j}}\left[1-2(-1)^{\delta_{jt}}\beta'e^{i\xi_{j}'}\right]|j\rangle,
\label{psiprimesecond}
\end{equation}
where $\beta' = \cos\frac{\mu_{0}}{2}|\beta|$ and $\xi_{j}' = \xi-\epsilon_{j}+\frac{\mu_{0}}{2}$. The bounds on $\beta,\xi,\mu_{0}$ and $\epsilon_{j}$ give
\begin{eqnarray}
(1-\beta')&\leq& 0.5\Delta_{t}^{2}+0.125\Delta_{0}^{2}+2\alpha_{U}^{2}, \nonumber \\
|\xi_{j}'| &\leq &2\Delta_{t}+0.5\Delta_{0}. \label{betaxiprimebound}
\end{eqnarray}

Using Eq. (\ref{psiprimesecond}), we get $|\langle j|\psi'\rangle /U_{j0}|_{j\in T}=|1+2\beta'e^{i\xi_{j}'}|$. The bounds on $\beta'$ and $\xi_{j}'$ then yield
\begin{equation}
\left(3 - \left|\frac{\langle j|\psi'\rangle}{U_{j0}}\right|\right)_{j\in T} \leq \frac{7}{3}\Delta_{t}^{2}+\frac{2}{3}\Delta_{t}\Delta_{0}+\frac{1}{3}\Delta_{0}^{2}+4\alpha_{U}^{2}. \label{ratiobound} 
\end{equation}

\textbf{Special Case:} Consider the situation $S_{0}=R_{0}^{\varphi_{0}}$, i.e. $S_{0}'=I$. In this case, $U_{1}|0\rangle =|\psi'\rangle$, and we have $\rho_{j} = |\langle j|\psi'\rangle/U_{j0}|$ which obeys the bound (\ref{ratiobound}) for $j\in T$. Using Eq. (\ref{kapparho}), we get  
\begin{equation}
(3-\kappa) \leq \frac{7}{3}\Delta_{t}^{2}+\frac{2}{3}\Delta_{t}\Delta_{0}+\frac{1}{3}\Delta_{0}^{2}+4\alpha_{U}^{2}.
\end{equation}
Thus the projection on the target subspace is amplified by a factor close to $3$ as $\Delta_{t},\Delta_{0}$ and $\alpha_{U}$ are small quantities. The main idea behind recursion is to note that the above analysis holds for any unitary operator $U$, and hence it also holds for $U_{1}$ which is a unitary operator. Therefore, $U_{2}=U_{1}S_{0}U_{1}^{\dagger}S_{t}U_{1}$ will obey 
\begin{equation}
(3-\kappa_{2}) = \left(3-\frac{\alpha_{U_{2}}}{\alpha_{U_{1}}}\right)\leq \frac{7}{3}\Delta_{t}^{2}+\frac{2}{3}\Delta_{t}\Delta_{0}+\frac{1}{3}\Delta_{0}^{2}+4\alpha_{U_{1}}^{2},
\end{equation}
where $\alpha_{U_{2}} = \sqrt{\sum_{j\in T}|\langle j|U_{2}|0\rangle|^{2}}$.
Thus the projection on the target subspace is amplified again by a factor close to $3$, making the total amplification close to $3^{2}=9$. Continuing the process, the $m^{th}$ level of recursion gives $\alpha_{U_{m}}=\prod_{l=1}^{m}\kappa_{l}\alpha_{U}$, where $(3-\kappa_{l})\leq \frac{7}{3}\Delta_{t}^{2}+\frac{2}{3}\Delta_{t}\Delta_{0}+\frac{1}{3}\Delta_{0}^{2}+4\alpha_{U_{l-1}}^{2}$. As long as $\alpha_{U_{m}}^{2}\ll 1$, the complete amplification factor obeys $3^{m}\geq\prod_{l=1}^{m}\kappa_{l}\geq \overline{\kappa}^{m}$, where
\begin{equation}
\overline{\kappa} \approx 3-\frac{7}{3}\Delta_{t}^{2}-\frac{2}{3}\Delta_{t}\Delta_{0}-\frac{1}{3}\Delta_{0}^{2}.
\end{equation} 

This analysis shows that $m$ levels of recursion can be used for amplifying the projection on target subspace to at least $\alpha_{U_{m}}=O(\overline{\kappa}^{m}\alpha_{U})$. We can always choose $m$ such that the condition $\alpha_{U_{m}}^{2}=c \ll 1$ is satisfied, and then repeat the algorithm $c^{-1}$ times to get a target state. The number of queries required by the algorithm to get a target state is, therefore, at most $q_{m}=O(3^{\log_{\overline{\kappa}}(1/\alpha_{U})})=O(1/\alpha_{U}^{\log_{\overline{\kappa}}3})$. In other words, the query complexity of the algorithm is $O(\alpha_{U}^{-(1+p)})$, with
\begin{equation}
0\leq p=\left(\frac{\log 3}{\log\overline{\kappa}}-1\right)\leq 0.71\Delta_{t}^{2}+0.20\Delta_{t}\Delta_{0}+0.10\Delta_{0}^{2}.
\end{equation}

\textbf{General Case}: For more general $S_{0}$ transformations, the state $U_{1}|0\rangle = US_{0}'U^{\dagger}|\psi'\rangle$ is not equal to $|\psi'\rangle$. $S_{0}'$ is close to identity, however, and $\|US_{0}'U^{\dagger}-I\|=\|S_{0}'-I\|=\Delta_{0}$. Upto a phase factor, we have
\begin{equation}
\langle j|US_{0}'U^{\dagger}=cos\gamma_{j}\langle j|+sin\gamma_{j}\langle x_{j}|, \label{xrangledefine}
\end{equation}
where $|x_{j}\rangle$ is a normalized vector orthogonal to $|j\rangle$. As $\|US_{0}'U^{\dagger}-I\|=\Delta_{0}$, we have the bound $|\gamma_{j}|\leq \Delta_{0}$ so that $sin\gamma_{j} \approx \gamma_{j}$. Now 
\begin{equation}
(U_{1})_{j0} = \langle j|US_{0}'U^{\dagger}|\psi'\rangle = cos\gamma_{j}\langle j|\psi'\rangle+\gamma_{j}\langle x_{j}|\psi'\rangle.
\end{equation}
Using Eq. (\ref{psiprimesecond}) for $|\psi'\rangle$, we find the ratio $\rho_{j}$ to be
\begin{equation}
\rho_{j\in T} = \left|\frac{(U_{1})_{j0}}{U_{j0}}\right|_{j\in T} =\left|cos\gamma_{j}e^{i\phi_{j}}(1+2\beta'e^{i\xi_{j}'})+\gamma_{j}\frac{\langle x_{j}|\psi'\rangle}{U_{j0}}\right| . \label{kappageneralfirst}
\end{equation}
As $|\psi'\rangle = UR_{0}^{\varphi_{0}}U^{\dagger}|\psi\rangle$, we have $|\langle\psi'|U|0\rangle| = |\langle\psi|U|0\rangle| = |\beta|$. Hence, up to a phase factor,
\begin{equation}
|\psi'\rangle = \beta U|0\rangle + \overline{\beta}|y\rangle\ ,\ \overline{\beta} =\sqrt{1-|\beta|^{2}},
\end{equation} 
where $|y\rangle$ is a normalised vector orthogonal to $U|0\rangle$. The bound on $\beta$ (\ref{betanonprimebound}) implies $\overline{\beta} \leq \sqrt{\Delta_{t}^{2}+4\alpha_{U}^{2}}$. Eq. (\ref{kappageneralfirst}) then reduces to
\begin{eqnarray}
\rho_{j\in T} &=&|C_{1j}+C_{2j}+C_{3j}|, \nonumber \\
C_{1j}&=& cos\gamma_{j}e^{i\phi_{j}}(1+2\beta'e^{i\xi_{j}'}), \nonumber \\
C_{2j}&=& \gamma_{j} \beta \frac{\langle x_{j}|U|0\rangle}{U_{j0}}, \nonumber \\
C_{3j}&=&\gamma_{j} \overline{\beta} \frac{\langle x_{j}|y\rangle}{U_{j0}}.  \label{kappageneralsecond} 
\end{eqnarray}

Since $cos\gamma_{j} = 1-O(\Delta_{0}^{2})$ and $1+2\beta'e^{i\xi_{j}'} = 3-O(\Delta_{t}^{2},\Delta_{0}^{2},\Delta_{t}\Delta_{0})$ (as proved earlier), we have $|C_{1j}| \approx 3$ for small $\{\Delta_{t},\Delta_{0}\}$. Using the definition (\ref{xrangledefine}) of $\langle x_{j}|$ and the bound $\gamma_{j} \leq \Delta_{0}$,
\begin{equation}
\langle x_{j}|U|0\rangle = U_{j0}\frac{1-cos\gamma_{j}}{\gamma_{j}} = U_{j0}O(\Delta_{0}), 
\end{equation}  
which makes $C_{2j} = O(\Delta_{0}^{2})$ and $|C_{1j}+C_{2j}| = 3-O(\Delta_{t}^{2},\Delta_{0}^{2},\Delta_{t}\Delta_{0})$. The ratio $\rho_{j\in T}$ will then be close to $3$ iff 
\begin{equation}
|C_{3j}| = \gamma_{j} \overline{\beta} \left|\frac{\langle x_{j}|y\rangle}{U_{j0}}\right| \ll 3.
\end{equation}
By their definitions, the vectors $|x_{j}\rangle$ and $|y\rangle$ depend upon the eigenvalues of $S_{0}$ and $S_{t}$ respectively. In most cases, the eigenvalues of these two different operators are uncorrelated (in case they are correlated, we need to randomize one of them by random operations), and hence $|x\rangle$ and $|y\rangle$ are two relatively random unit vectors in the $N$-dimensional Hilbert space. So the expectation value of their inner product $|\langle x|y\rangle|$ is $1/\sqrt{N}$, and the above condition translates to
\begin{equation}
\frac{\gamma_{j}\overline{\beta}}{\sqrt{N}} \ll 3|U_{j0}|, \label{recursivecondition}
\end{equation}   
As long as this condition is satisfied for all $j\in T$, the ratio $\rho_{j\in T}$ and the amplification factor $\kappa$ are close to $3$. More precisely,
\begin{equation}
\kappa = 3-O(\Delta_{t}^{2},\Delta_{0}^{2},\Delta_{t}\Delta_{0}).
\end{equation}  
If this condition is not satisfied for a particular target state $j$, then the amplitude along it will not be amplified by the recursive algorithm as if it were a non-target state. 

The condition (\ref{recursivecondition}) is only a sufficient, not necessary, condition for $\kappa$ to be close to $3$. If it is satisfied for the first level of recursion then it is automatically satisfied for higher levels as $|U_{j0}|<|(U_{l})_{j0}|$ for any $l$. Also, even if this condition is not satisfied then amplification may still be possible by a factor greater than $1$, but not close to $3$. Note that if $\gamma_{j}$ or $\overline{\beta}$ is $O(U_{j0})$ then the condition is satisfied. It can be shown that this is the case when either of $S_{t}$ or $S_{0}$ becomes a selective phase-rotation $R_{t}$ or $R_{0}$ (the special case discussed earlier corresponds to $S_{0}=R_{0}$). Also, the condition is always satisfied for $U=W$ as $W_{j0} = 1/\sqrt{N}$ and $\gamma_{j}\overline{\beta} \leq \Delta_{0}\sqrt{\Delta_{t}^{2}+4\alpha_{U}^{2}}\ll 1$.

\subsection{Comparison with Grover's algorithm} 

When $\{S_{t},S_{0}\}=\{I_{t},I_{0}\}$, the recursive algorithm reduces to the iterative Grover's algorithm and the optimal query complexity of $O(1/\alpha_{U})$ is achieved. The state at $m^{th}$ level of recursion $U_{m}|0\rangle$ is nothing but $(3^{m}-1)/2$ applications of $UI_{0}U^{\dagger}I_{t}$ on the initial state $U|0\rangle$. Explicitly, with $I_{t}^{\dagger}=I_{t}$ and $I_{0}^{\dagger}=I_{0}$,
\begin{eqnarray*}
U_{m+1} &=&(UI_{0}U^{\dagger}I_{t})^{q_{m}}UI_{0}U^{\dagger}(I_{t}^{\dagger}UI_{0}^{\dagger}U^{\dagger})^{q_{m}}I_{t}(UI_{0}U^{\dagger}I_{t})^{q_{m}}U \\
        &=&(UI_{0}U^{\dagger}I_{t})^{q_{m}}UI_{0}U^{\dagger}(I_{t}UI_{0}U^{\dagger})^{q_{m}}I_{t}(UI_{0}U^{\dagger}I_{t})^{q_{m}}U \\
        &=& (UI_{0}U^{\dagger}I_{t})^{3q_{m}+1}U.
\end{eqnarray*}
With $q_{0}=0$, $U_{m}=(UI_{0}U^{\dagger}I_{t})^{(3^{m}-1)/2}U$ is just quantum amplitude amplification, except for the jumps in the number of queries.

In recursive quantum search, we are interested in the amplification factor $\kappa = \alpha_{U_{1}}/\alpha_{U}$ of the projection on the target subspace, achieved by applying $US_{0}U^{\dagger}S_{t}$ to $U|0\rangle$. Detailed eigenspectrum of $US_{0}U^{\dagger}S_{t}$ is not of much relevance, since  what matters is only one (rather than multiple) application of $US_{0}U^{\dagger}S_{t}$. In general, the state $US_{0}U^{\dagger}S_{t}U|0\rangle= UI_{0}U^{\dagger}I_{t}U|0\rangle + |\Delta\rangle$, where $|\Delta\rangle$ has norm $O(\Delta_{t},\Delta_{0})$. $\kappa$ is certainly close to $3$, when $\{\Delta_{t},\Delta_{0}\}\ll O(\alpha_{U})$. What we have shown above is that even when $\{\Delta_{t},\Delta_{0}\} \not\ll O(\alpha_{U})$, $\kappa$ can be close to $3$. That is because what matters for $\kappa$ is not the norm of $|\Delta\rangle$ but its projection on the target subspace, which can be small compared to $\alpha_{U}$ even when $\{\Delta_{t},\Delta_{0}\}\not\ll O(\alpha_{U})$.

The recursive algorithm needs $O(1/\alpha_{U}^{1+O(\Delta^{2})})$ queries, with $\Delta = O(\Delta_{t},\Delta_{0})$ characterizing the size of errors. The increase in query complexity, due to nonzero $\Delta$, is only a constant factor provided $\Delta = O(\sqrt{-1/\log\alpha_{U}})$. This is a much better performance than the quantum amplitude amplification algorithm which needs $\Delta = O(1/\alpha_{U})$ for success. Furthermore, the recursive algorithm can succeed even for larger $\Delta$ at the cost of more queries.

\section{Discussion}

Finally we consider the situation when $\{I_{0},I_{t}\}$ are replaced by non-diagonal operators $\{P,Q\}$. The iterative algorithm then evaluates $(UP^{\dagger}U^{\dagger}Q^{\dagger}UPU^{\dagger}Q)^{n} U|0\rangle$. Using diagonal decompositions of $\{P,Q\}$, i.e. $P=E_{P}S_{0}E_{P}^{\dagger}$ and $Q=E_{Q}S_{t}E_{Q}^{\dagger}$ with $S_{0}$ and $S_{t}$ diagonal, that becomes $E_{Q}(VS_{0}^{\dagger}V^{\dagger}S_{t}^{\dagger}VS_{0}V^{\dagger}S_{t})^{n} VE_{P}^{\dagger}|0\rangle$, where $V=E_{Q}^{\dagger}UE_{P}$. The algorithm therefore converges to the target state in $O(1/\alpha_{V})$ steps, provided $\{S_{t},S_{0}\}$ satisfy conditions for successful quantum search and $(E_{P})_{00},(E_{Q})_{tt}$ are close to $1$. The condition $(E_{P})_{00},(E_{Q})_{tt} \approx 1$ is important for any search algorithm, because only then we can rightfully call the transformations \emph{selective}, performing an operation on the intended state and leaving the other states alone. Thus, as long as $V_{t0}\not\ll U_{t0}$, there is no significant slowdown in quantum search.  

Similarly, the recursive algorithm evaluates $U_{m}|0\rangle = E_{Q}V_{m}E_{P}^{\dagger}|0\rangle$ at the $m^{th}$ level, with
\begin{equation}
V_{m} = V_{m-1}S_{0}V_{m-1}^{\dagger}S_{t}V_{m-1}.
\end{equation}
As before, the algorithm succeeds, provided $\{S_{t},S_{0}\}$ satisfy conditions for successful quantum search and $(E_{P})_{00},(E_{Q})_{tt}$ are close to $1$.

Next we point out a few applications of our algorithms.

\textbf{(1) Correction of Certain Systematic Errors:}
Quantum amplitude amplification is a repetitive application of the operator $\overline{\mathcal{G}}=UI_{0}U^{\dagger}I_{t}$. Small errors in $\overline{\mathcal{G}}$ may accumulate over iterations to produce a large deviation at the end, causing the algorithm to fail. Completely random errors have to be protected against, using the techniques of quantum error correction and fault-tolerant quantum computation~\cite{qec}. That adds redundancy to the quantum states and gates, i.e. extra resources, to overcome small errors. For errors exhibiting specific structures, however, it is worthwhile to investigate whether the dependence on quantum error-correction can be reduced by designing quantum algorithms that are intrinsically robust to these errors.

In this paper, we have studied a particular class of systematic errors, those that are perfectly \emph{reproducible} and \emph{reversible}. For an imperfect apparatus in this category, we have presented two algorithms that exploit the structure of errors and succeed in quantum search while the standard quantum search fails. These type of errors are not uncommon, e.g. the errors arising from imperfect pulse calibration and offset effect in NMR systems~\cite{avik}. Thus our algorithms offer a significant flexibility in physical implementation of quantum search.

\textbf{(2) Handling Errors in Workspace:}
The $I_{t}$ transformation used in quantum search is implemented using an oracle. A typical implementation uses an ancilla qubit initialized to the $\frac{|0\rangle-|1\rangle}{\sqrt{2}}$ state, and a C-NOT gate applied to it from a Boolean function $f(j)$. In general, $f(j)$ has to be computed using the techniques of reversible computation, and has to be uncomputed afterwards to ensure reversibility. Inevitably, we need to couple our search-space to an ancilla workspace to implement $I_{t}$, and the two get entangled. For a perfect algorithm, the workspace returns to its initial state at the end of the algorithm, and the search-space and the workspace get disentangled. But when there are errors, the workspace may not exactly return to its initial state, leaving some entanglement between the search-space and the workspace at the end. That deteriorates the performance of quantum search, and our algorithms come to rescue in such cases.

Let $\widehat{\mathcal{H}}=\mathcal{H}_{s}\otimes \mathcal{H}_{w}$ be the joint Hilbert space of the search-space and the workspace. The perfect oracle is $I_{t}= \sum_{j}\left(|j\rangle\langle j|_{f(j)=1}\otimes (-I)+|j\rangle\langle j|_{f(j)=0}\otimes I\right)$. In case of imperfect oracles, it may become $Q= \sum_{j}\left(|j\rangle\langle j|_{f(j)=1}\otimes A+|j\rangle\langle j|_{f(j)=0}\otimes B\right)$, where $A,B$ are unitary operators. First consider the case $B=I$, i.e. the workspace remains unaltered for $f(j)=0$. With the diagonal decomposition $A=E_{Q}S_{t}E_{Q}^{\dagger}$, we have $Q= \sum_{j}\big(|j\rangle\langle j|_{f(j)=1}\otimes E_{Q}S_{t}E_{Q}^{\dagger} +|j\rangle\langle j|_{f(j)=0}\otimes I\big)$. That is equivalent to $R_{t}$ of section II, performing a selective phase-rotation by $\phi_{k}$ of the effective target state $|j\rangle_{f(j)=1}\otimes |E_{Q}(\phi_{k})\rangle$ in $\widehat{\mathcal{H}}$, where $|E_{Q}(\phi_{k})\rangle$ is the eigenvector of $A$ with the eigenvalue $e^{i\phi_{k}}$. Our iterative algorithm would use $\widehat{\mathcal{T}}=\widehat{U}I_{\hat{0}}\widehat{U}^{\dagger}S_{t}^{\dagger}\widehat{U}I_{\hat{0}}\widehat{U}^{\dagger}S_{t}$, where $\widehat{U}=U\otimes I$ and $I_{\hat{0}}$ is the selective phase-inversion of $|\hat{0}\rangle = |0\rangle\otimes|0_{w}\rangle$ with $|0_{w}\rangle$ the initial state of the workspace. As shown in section II, iterating $\widehat{\mathcal{T}}$ leads us to a state $|j\rangle_{f(j)=1}\otimes|\psi\rangle$, whose projection on the search-space is a target state. The number of queries depends on the eigenvalues of $A$, but it will be $O(1/\alpha_{U})$ as long as the eigenvalues are away from $1$. The same result applies if the operator $A$ is different for different target states. Note that this is a much relaxed criterion than the phase-matching condition which demands the eigenvalues of $A$ to be within $O(\alpha_{U})$ of $-1$.

When $B \neq I$ as well, the iterative algorithm cannot take us to a target state and we have to use the recursive algorithm. The condition that $\|S_{t}-I_{t}\|$ should be small, restricts $A$ to be close to $-I$ (unlike the iterative algorithm, which allows a much wider range of $A$) and $B$ to be close to the identity operator. For small errors in the workspace transformations, therefore, quantum search works and complete elimination of the entanglement between the search-space and the workspace is not necessary.

\textbf{(3) Bounded Error Quantum Search:}
Our recursive search algorithm is similar to the quantum search algorithm on bounded error inputs by Hoyer {\it et al.}~\cite{hoyerbound} (labeled HMW henceforth), except that our error model is much more general. HMW considered \emph{computationally} imperfect oracles, which provide the correct value of $f(j)$ not with certainty but with a probability close to $1$. For instance, if $j$ is a target (non-target) state, the Boolean oracle may output $1$ ($0$) with at least a probability $9/10$. We have considered \emph{physically} imperfect oracles, where the errors affect the unitary transformations corresponding to the oracle. In particular, the algorithm by HMW (see facts $1,2$ in section $3$ of \cite{hoyerbound}) uses fixed unitary transformations $(S_{0})_{\rm hmw},(S_{1})_{\rm hmw}$ (amplitude amplification) and $E_{\rm hmw}$ (error reduction), with $(S_{1})_{\rm hmw}$ replacing the oracle $I_{t}$. Our algorithm applies to the situation where these unitary transformations themselves contain errors. We have shown that as long as the errors are small, quantum search is possible.

Indeed, the HMW error model can be reduced to our error model. The HMW oracle transformation $O$ computes the value of $f(j)$ using workspace qubits and stores it in a qubit. It takes the initial state $\sum_{j}a_{j}|j\rangle|0_{w}\rangle|0\rangle$ to $\sum_{j}a_{j}|j\rangle(\sqrt{p_{j}}|\psi_{j1}\rangle|1\rangle+\sqrt{1-p_{j}}|\psi_{j0}\rangle|0\rangle)$, where $|\psi_{jb}\rangle,\ b\in\{0,1\}$ denote the workspace states. The probability $p_{j}$ is at least $9/10$ if $f(j)=1$ and at most $1/10$ if $f(j)=0$. Consider the operator $\overline{\mathcal{G}}_{O}=OI_{0}O^{\dagger}S_{1}O$ instead of only $O$, where $S_{1}$ inverts the states with last qubit $|1\rangle$ and $I_{0}$ is the selective phase-inversion of the $|0_{w}\rangle|0\rangle$ state. The operator $\overline{\mathcal{G}}_{O}$ is an amplitude amplification operator, and its eigenvalues are $e^{\pm 2i\theta_{j}}$ with $sin^{2}\theta_{j}=p_{j}$~\cite{qaa}. Hence for $f(j)=1(0)$, the eigenvalues are close to $-1(1)$. This is similar to the workspace error model discussed above, where $\|S_{t}-I_{t}\|$ is small.

Moreover, if we assume that there are no errors in workspace transformations, our error model can also be reduced to the HMW error model. We simply attach a qubit to the workspace in the $\frac{|0\rangle+|1\rangle}{\sqrt{2}}$ state. A controlled $S_{t}$ transformation takes the qubit to the state $(|0\rangle+e^{i\phi_{j}}|1\rangle)/\sqrt{2}$, where $e^{i\phi_{j}}$ are eigenvalues of $S_{t}$. A Hadamard gate then transforms the qubit to the state $e^{i\phi_{j}/2}\big(cos\frac{\phi_{j}}{2}|0\rangle-isin\frac{\phi_{j}}{2}|1\rangle\big)$. Since $\phi_{j}=\pi f(j)+\epsilon_{j}$ with small $|\epsilon_{j}|$, we obtain the HMW model. 

The difference arises when the workspace transformations of the HMW model also suffer from errors. To get rid of these errors, we cannot keep on attaching extra ancilla qubits till the new ancilla qubits are free of errors. Our results show that there is no need to worry about it, and recursion works as long as the errors are small. 

A peculiar feature of the HMW error model is that the imperfect oracle can be used to simulate an almost perfect oracle by making $O(\log{N})$ oracle queries. Thereafter, the standard quantum search can be used. In our model, we cannot simulate $I_{t}$ using $S_{t}$. In fact, we have shown that there is no need to simulate $I_{t}$; $S_{t}$ is good enough for quantum search as long as it is close to $I_{t}$. More importantly, our algorithm also works when $I_{0}$ is affected by errors, a case not considered by HMW.       
 
To conclude, we have presented two algorithms which allow a significant flexibility in the selective transformations used by quantum search. The iterative algorithm takes $O(\sqrt{N})$ queries and requires the oracle to be neutral for non-target states. But the oracle may mark the target states by phases other than phase-inversion, and hence almost any oracle transformation is good enough for quantum search. The recursive algorithm tackles the situations when the oracle perturbs non-target states also. For error size $\Delta$, it reaches a target state using $O(\sqrt{N}\cdot N^{O(\Delta^{2})})$ queries. Needless to say, errors are inevitable in any physical implementation of quantum search. As long as the errors are small, the algorithms we have constructed are more robust and better adapted to physical implementation than the standard quantum search.

\textbf{Acknowledgements}: I thank Prof. Apoorva Patel for going through the manuscript and for useful comments and discussions. I thank Avik Mitra and Prof. Anil Kumar for discussions on the experimental implementation of the iterative algorithm.


\begin{thebibliography}{9}
\bibitem{grover1} L.K. Grover, Phys. Rev. Lett. \textbf{79}, 325 (1997). 
\bibitem{grover2} L.K. Grover, Phys. Rev. Lett. \textbf{80}, 4329 (1998).
\bibitem{qaa} G. Brassard, P. Hoyer, M. Mosca, and A. Tapp, Contemporary Mathematics (American Mathematical Society, Providence), \textbf{305}, 53 (2002). e-print quant-ph/0005055.
\bibitem{long1} G.L. Long, Y.S. Li, W.L. Zhang, and L. Niu, Phys. Lett. A \textbf{262}, 27 (1999).
\bibitem{hoyer1} P. Hoyer, Phys. Rev. A \textbf{62}, 052304 (2000).
\bibitem{long2} G.L. Long, Y.S. Li, W.L. Zhang, and C.C. Tu, Phys. Rev. A \textbf{61}, 042305 (2000).
\bibitem{long3} G.L. Long, C.C Tu, Y.S. Li, W.L. Zhang, and H.Y. Yan, e-print quant-ph/9911004.
\bibitem{avik} A. Mitra, A. Tulsi, and A. Kumar, \textit{Experimental NMR Implementation of error-resistant quantum search algorithm}, submitted.
\bibitem{tulsi}A. Tulsi, \textit{A general framework for quantum search algorithms}, manuscript in preparation. 
\bibitem{farhi} E. Farhi and S. Gutmann, Phys. Rev. A \textbf{57}, 2403 (1998).
\bibitem{fenner} S.A. Fenner, e-print quant-ph/0004091. 
\bibitem{baekwon} J. Bae and Y. Kwon, Phys. Rev. A \textbf{66}, 012314 (2000).
\bibitem{roland} J. Roland and N.J. Cerf, Phys. Rev. A \textbf{68}, 062311 (2003). 
\bibitem{hoyerbound} P. Hoyer, M. Mosca, and R.D. Wolf, Proc. ICALP 03 (2003), e-print quant-ph/0304052.
\bibitem{private} L.K. Grover, private communication.
\bibitem{qec} J. Preskill, Proc. R. Soc. London, Ser. A, \textbf{454}, 385 (1998). 
\end{thebibliography}
\end{document}